\documentclass[12pt]{article}%
\usepackage{cite}
\usepackage{amsmath}
\usepackage{amsfonts}
\usepackage{amssymb}
\usepackage{graphicx}%
\providecommand{\U}[1]{\protect\rule{.1in}{.1in}}
\begin{document}

\title{On Kinetic Theory of Energy Losses in Randomly Heterogeneous Medium}
\author{Sergey Panyukov, Andrei Leonidov\\\; \\Theoretical Physics Department \\P.N. Lebedev Physics Institute, Moscow, Russia}
\maketitle

\begin{abstract}
We derive equation describing distribution of energy losses of the particle
propagating in fractal medium with quenched and dynamic heterogeneities. We
show that in the case of the medium with fractal dimension $2<D<3$ the losses
$\Delta$ are characterized by the sublinear anomalous dependence $\Delta\sim
x^{\alpha}$ with power-law dependence on the distance $x$ from the surface and
exponent $\alpha=D-2$.

\end{abstract}

This letter is devoted to studying statistical properties of the collisional
energy losses suffered by a high energy particle passing through a randomly
inhomogeneous disordered medium. Exploration of the properties of complex
media with the help of test particles propagating through it is one of the
most important scientific instruments used in physics allowing, in particular,
to study a response of the medium to particle beams or radiation coming
through the medium under study. A range of possible applications is quite
broad, from the physics of high energy collisions to polymer physics, to name
the few. In all cases the main quantity under study is the distribution of
energy losses of the test particle studied as a function of the distance
covered by the test particle in the medium.

The original context of the problem was related to examination of ionization
losses of high energy particles in ordinary homogeneous amorphous matter
(energy straggling), see e.g. \cite{Book}. The microscopic picture underlying
the energy losses in this case was, evidently, that of a series of inelastic
collisions of the projectile with atomic electrons resulting in
excitation/ionization of the corresponding atoms. In between the scattering
events the projectile trajectory is ballistic and its energy does not change.
In the continuum limit the problem can be reformulated, for a high energy
projectile and energy losses small compared to the energy of the incident
particle, in terms of a one-dimensional kinetic equation suggested by Landau
\cite{L44}, in which the role of time is played by the distance along the
straight line trajectory covered by the projectile in the medium.

The focus of the present paper is on the statistical properties of energy
straggling in matter characterized by strong random inhomogeneities with the
special emphasis on the case of fractal medium. Our consideration of the
microscopic energy loss model in a strongly inhomogeneous random medium with
power-like correlations is realized at the level analogous to the Continuous
Random Walk (CTRW)\cite{CTRW_model} that was, in particular, actively explored
to describe thermal diffusion of particles in a fractal
medium.\cite{KBZ84,S84,BKWZ84,IV03} Physically, the main difference between
the stochastic process describing the energy loss and the CTRW is in straight
line trajectory of high energy particle, while CTRW describes haotic
trajectories of particle induced by random Brownian forces. From mathematical
point of view the distinction of these processes is that the energy loss
problem corresponds to a sum of random positive quantities -- energy losses at
scattering events, whereas random particle displacements in CTRW can have any sign.

Let us formulate the basic microscopic model of energy loss studied in this
paper and consider a high energy particle incident on the medium containing
randomly placed scattering centers. In the high energy approximation the
trajectory of the projectile is a straight line and the particle is assumed to
interact with all the scattering centers that happen to lie at the
projectile's trajectory. More precisely, let us consider the particle with
high energy $E_{0}$ entering the medium at the point $x=0$. Our main goal is
to compute the distribution $f\left(  \Delta,x\right)  $ of its energy loss
$\Delta$ at some depth $x$ from the surface.

\begin{figure}[tbh]
\begin{center}
\includegraphics[
height=0.5902in, width=3.5826in
]{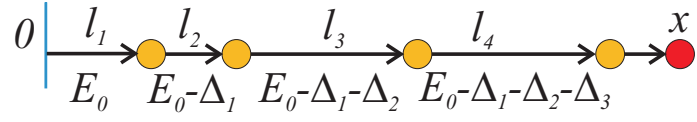}
\end{center}
\caption{Particle propagating in the medium looses the energy $\Delta
=\Delta_{1}+\Delta_{2}+\cdots+\Delta_{n}$ in $n$ scattering events taking
place at points separated by intervals $\{l_{i}\}$ along the trajectory of the
projectile.}%
\label{toy}%
\end{figure}

The energy loss $\Delta$ at the point $x$ in a given event is fully
characterized by the set of energy losses at each of scattering event
$\{\Delta_{i}\}$, $i=1,\cdots,n$ so that $\Delta=\Delta_{1}+\Delta_{2}%
+\cdots+\Delta_{n}$. The corresponding configuration of the scattering centers
is, in turn, fully specified by the set of distances $l_{1},l_{2},\cdots
,l_{n},l_{n+1}$, where $l_{1}$ is the distance between the surface and the
first scattering center, $\{l_{i}\}$ are the distances between the $i$'th and
$i+1$'th scattering centers and, finally, $l_{n+1}$ is the distance between
the last scattering center and the observation point $x$ ($x=l_{1}%
+l_{2}+\cdots+l_{n+1}$), see Fig.~\ref{toy}.

In the assumption of independent energy losses at different scattering centers
with the distribution of energy losses $w(\varepsilon)$ the probability
distribution $f(\Delta\mathbf{,}x)$ of the cumulative energy loss $\Delta$ at
some point $x$ is fully described by the probability densities $\psi(l)$ of
having a spatial distance $l$ between two scattering centers, see
Fig.~\ref{toy}. For large number of scattering events the energy losses do not
depend on the probability density $\varphi(l_{1})$ of the first event, and in
follows we will simply assume that $\varphi(l_{1})=\psi(l_{1})$. The problem
of random loss is similar to that of the usual random walk, with the distance
$x$ and local energy loss $\Delta$ replacing the time and displacement
correspondingly. From this analogy the energy loss distribution function
$f\left(  \Delta\mathbf{,}x\right)  $ is described by CTRW-like equation:
\begin{equation}
f\left(  \Delta\mathbf{,}x\right)  =\delta\left(  \Delta\right)  \Psi\left(
x\right)  +\int_{0}^{x}dx^{\prime}\psi\left(  x-x^{\prime}\right)  \int
_{0}^{\Delta}d\varepsilon w\left(  \varepsilon\right)  f\left(  \Delta
-\varepsilon\mathbf{,}x^{\prime}\right)  ,\label{fDx}%
\end{equation}
where $\Psi\left(  x\right)  =\int_{x}^{\infty}dy\psi(y)$ is the probability
of having no scattering events in the interval $[0,x]$. Eq.~(\ref{fDx}) is
conveniently solved by using the double Laplace transform
\[
\tilde{f}\left(  p,q\right)  \equiv\int_{0}^{\infty}d\Delta e^{-p\Delta}%
\int_{0}^{\infty}dxe^{-qx}f\left(  \Delta,x\right)
\]
Introducing the finction $\tilde{g}\left(  q\right)  \equiv\tilde{\psi}\left(
q\right)  /[1-\tilde{\psi}\left(  q\right)  ]$ the corresponding equation for
$\tilde{f}\left(  p,q\right)  $ can be written in the form
\begin{equation}
\tilde{f}\left(  p,q\right)  =1/q+\tilde{g}\left(  q\right)  \left[  \tilde
{w}\left(  p\right)  -1\right]  \tilde{f}\left(  p,q\right)  \label{ff0}%
\end{equation}
which, in turn, corresponds to the following version of the original kinetic
equation (\ref{fDx}):
\begin{equation}
f\left(  \Delta\mathbf{,}x\right)  =\delta\left(  \Delta\right)  +{\int
_{0}^{x}}dx^{\prime}g\left(  x-x^{\prime}\right)  {\int_{0}^{\infty}%
}d\varepsilon w\left(  \varepsilon\right)  \left[  f\left(  \Delta
-\varepsilon,x^{\prime}\right)  -f\left(  \Delta,x^{\prime}\right)  \right]
\label{ff1}%
\end{equation}
The function\textbf{\ }$g\left(  r\right)  $\textbf{\ }is found by inverse
Laplace transform of the function $\tilde{g}\left(  q\right)  $ and it has the
meaning of the average density of scatterings along the direction of particle
propagation at the distance $r$ from the last scattering.

In general, the function $g\left(  r\right)  $ depends on characteristics of
the medium, and it can be related to the so-called structure function of the
medium
\begin{equation}
G\left(  \mathbf{r}\right)  =\left\langle \sum\nolimits_{n\neq0}\delta\left(
\mathbf{x}_{i}-\mathbf{x}_{i+n}-\mathbf{r}\right)  \right\rangle ,\label{SF}%
\end{equation}
where $\mathbf{x}_{i}$ are coordinates of the $i$-th scattering center, by
\begin{equation}
g\left(  r\right)  =a^{2}G\left(  r\right)  ,\label{gG}%
\end{equation}
$a^{2}$ is the scattering area of the particle. In many important cases of the
scattering of particles in a complex heterogeneous medium a microstructure of
the medium remains unknown, but its structure function $G\left(  r\right)  $
can be directly measured experimentally. In these cases the kinetic
equation~(\ref{ff1}) can be used to predict the spectrum of energy losses in
such a medium. The knowledge of the spectrum is extremely important, for
example, in the problem of radiation damage of the medium which is determined
not only by total adsorbed energy but also by the shape of the distribution of
energy losses.

Eq. (\ref{ff1}) may be considered as a generalization of the Landau equation
for ionization losses in amorphous media\cite{Book}, which can be written in
integral form:
\begin{equation}
f\left(  \Delta\mathbf{,}x\right)  =\delta(\Delta)+\frac{1}{a}\int_{0}%
^{x}dx^{\prime}{\int_{0}^{\infty}}d\varepsilon w\left(  \varepsilon\right)
\left[  f\left(  \Delta-\varepsilon,x^{\prime}\right)  -f\left(
\Delta,x^{\prime}\right)  \right]  \label{bk0}%
\end{equation}
This equation is obtained from Eq.~(\ref{ff1}) in the case of the constant
linear density of scattering centers $g\left(  x-x^{\prime}\right)  =1/a$.

Analytical solution of generalized kinetic equation~(\ref{ff1}) can be found
in the case of scattering medium with fractal dimension $D$, when the Laplace
transform of the function $g\left(  r\right)  $ has the form:%
\begin{equation}
\tilde{g}\left(  q\right)  =\left(  aq\right)  ^{-\alpha},\quad\alpha=D-2
\label{gq}%
\end{equation}
In the infrared limit $x\gg a$ one can approximate the Laplacw transform of
the function $w\left(  \varepsilon\right)  $ by $\tilde{w}\left(  p\right)
\simeq1-p{\bar{\varepsilon}}$, where ${\bar{\varepsilon}}=\int\varepsilon
w(\varepsilon)d\varepsilon$ is the average energy loss in a scattering event.
Calculating the inverse Laplace transform of Eq.~(\ref{ff0}), we get
\begin{equation}
f\left(  \Delta,x\right)  =\frac{1}{\overline{\varepsilon}}\left(  \frac{a}%
{x}\right)  ^{\alpha}W_{\alpha}\left[  \frac{\Delta}{\overline{\varepsilon}%
}\left(  \frac{a}{x}\right)  ^{\alpha}\right]  , \label{fdx}%
\end{equation}
where $W_{\alpha}$ is the Wright function\cite{GLM99}%
\begin{equation}
W_{\alpha}\left(  z\right)  =\sum\nolimits_{l=0}^{\infty}\frac{(-z)^{l}%
}{l!\Gamma\left(  1-\alpha-\alpha l\right)  } \label{Wa}%
\end{equation}
Using the distribution function (\ref{fdx}) one can compute the average energy
loss at some depth $x$:
\begin{equation}
\left\langle \Delta\left(  x\right)  \right\rangle =[\overline{\varepsilon
}/\Gamma\left(  1+\alpha\right)  ]\left(  x/a\right)  ^{\alpha},\qquad
0<\alpha<1 \label{Dav}%
\end{equation}
We conclude that in the case of the fractal medium the average energy loss is
characterized by fractional sublinear dependence on the distance. The
equations (\ref{fdx},\ref{Dav}) constitute the main result of the paper.

Studied in this paper problem of random energy loss can be considered as
\textquotedblleft dual\textquotedblright\ to the problem of thermal random
walk motion of low energy particles. The thermal diffusion of particles in a
fractal medium is known to be anomalously slow, see e.g. \cite{wang00}. The
reason of the subdiffusional motion in fractal porous medium is that a
particle is trapped in dead end pores and bottlenecks, so that diffusion is
slowed down and becomes anomalous. The physics of anomalous random energy loss
in the fractal medium is different and is related with the presence of
long-range correlations in positions of scattering centers. Approach developed
in this work can also be applied to describe particle propagation in the
system with dynamic heterogeneities formed at the critical point of phase transition.

\end{document}